\documentclass{aa}  
\pdfoutput=1
\usepackage{twoopt}

\bibpunct{(}{)}{;}{a}{}{,}             
  \newcommandtwoopt{\citeads}[3][][]{\href{http://adsabs.harvard.edu/abs/#3}%
    {\def\hyper@linkstart##1##2{}%
     \let\hyper@linkend\@empty\citealp[#1][#2]{#3}}}
  \newcommandtwoopt{\citepads}[3][][]{\href{http://adsabs.harvard.edu/abs/#3}%
    {\def\hyper@linkstart##1##2{}%
     \let\hyper@linkend\@empty\citep[#1][#2]{#3}}}
  \newcommandtwoopt{\citetads}[3][][]{\href{http://adsabs.harvard.edu/abs/#3}%
    {\def\hyper@linkstart##1##2{}%
     \let\hyper@linkend\@empty\citet[#1][#2]{#3}}}
  \newcommandtwoopt{\citeyearads}[3][][]%
    {\href{http://adsabs.harvard.edu/abs/#3}
    {\def\hyper@linkstart##1##2{}%
     \let\hyper@linkend\@empty\citeyear[#1][#2]{#3}}}
\makeatother

\usepackage{graphicx}
\usepackage{footnote}
\usepackage{amsmath}
\usepackage{amssymb}
\usepackage{xcolor}
\usepackage{pdflscape}
\usepackage{url}
\usepackage[varg]{txfonts}
\usepackage{bm}         
\usepackage{booktabs}
\usepackage{txfonts}
\usepackage[pagewise]{lineno}

\DeclareGraphicsExtensions{.pdf,.png,.jpg}

\begin{document}

   \title{ALMACAL. XII. Data characterisation and products}


   \author{Victoria Bollo,
\inst{1\thanks{Email:victoria.bollo@eso.org}}
          Martin Zwaan, \inst{1}
          C\'eline P\'eroux, \inst{1, 2}
          Aleksandra Hamanowicz, \inst{3}
          Jianhang Chen, \inst{4}
          Simon Weng, \inst{5, 6, 7}\\
          Rob\,J.~Ivison, \inst{1, 8, 9, 10}
          \and
          Andrew Biggs\inst{11}
          }

   \institute{European Southern Observatory, Karl-Schwarzschild-Str. 2, 85748 Garching near Munich, Germany \label{1}
              \and
             Aix Marseille Univ., CNRS, LAM, (Laboratoire d’Astrophysique de Marseille), UMR 7326, F-13388 Marseille, France \label{2}
            \and
            Space Telescope Science Institute, 3700 San Martin Drive, Baltimore, MD 21218, USA \label{3}
            \and
            Max-Planck-Institut für Extraterrestrische Physik (MPE), Giessenbachstrasse 1, D–85748 Garching, Germany \label{4}
            \and
            Sydney Institute for Astronomy, School of Physics A28, University of Sydney, NSW 2006, Australia \label{5}
            \and
            ARC Centre of Excellence for All Sky Astrophysics in 3 Dimensions (ASTRO 3D) \label{6}
            \and
            ATNF, CSIRO Space and Astronomy, PO Box 76, Epping, NSW 1710, Australia \label{7}
            \and 
            Institute for Astronomy, University of Edinburgh, Royal Observatory, Blackford Hill, Edinburgh EH9 3HJ, UK
            \label{8}
            \and
            School of Cosmic Physics, Dublin Institute for Advanced Studies, 31 Fitzwilliam Place, Dublin D02 XF86, Ireland
            \label{9}
            \and
            ARC Centre of Excellence for All Sky Astrophysics in 3 Dimensions (ASTRO 3D)
            \label{10}
            \and
            UK Astronomy Technology Centre, Royal Observatory, Blackford Hill, Edinburgh EH9 3HJ, UK \label{11}}

   \date{Received XXX; accepted XXX}

\titlerunning{ALMACAL XII --- Data products}
\authorrunning{V. Bollo et al.}
 
\abstract{
The ALMACAL survey is based on a database of reprocessed ALMA calibration scans suitable for scientific analysis, observed as part of regular PI observations.
We present all the data accumulated from the start of ALMA operations until May 2022 for 1047 calibrator fields across the southern sky spanning ALMA Bands 3 to 10 ($\sim 84 - 950$ GHz), so-called ALMACAL$-22$. 
Encompassing over 1000 square arcmin and accumulating over 2000 hours of integration time, ALMACAL is not only one of the largest ALMA surveys to date, but it continues to grow with each new scientific observation.
We outline the methods for processing and imaging a subset of the highest-quality data ('pruned sample').
Using deconvolution techniques within the visibility data (uv plane), we created data cubes as the final product for further scientific analysis.
We describe the properties and shortcomings of ALMACAL and compare its area and sensitivity with other sub-millimetre surveys. 
Notably, ALMACAL overcomes limitations of previous sub-millimetre surveys, such as small sky coverage and the effects of cosmic variance.
Moreover, we discuss the improvements introduced by the latest version of this dataset that will enhance our understanding of dusty star-forming galaxies, extragalactic absorption lines, active galactic nucleus physics, and ultimately the evolution of molecular gas.
}


   \maketitle
%

\section{Introduction} \label{sec:introduction}

Millimetre (mm) and sub-millimetre (sub-mm) observations offer a unique window into various astrophysical processes and phenomena in the Universe. 
At these wavelengths, key emission lines such as carbon monoxide (CO) and atomic and ionised carbon ([CI], [CII]) emit radiation that reveals essential information about molecular gas content, star-formation activity, and gas dynamics \citep{carilliCoolGasHighRedshift2013, tacconiEvolutionStarFormingInterstellar2020}. 
Sub-mm observations help us explore the cosmic baryon cycle and the complex interplay between cold and dense molecular gas reservoirs, star formation activity, ionised gas, and dust emission \citep{perouxCosmicBaryonMetal2020}.

Modern interferometers have produced large datasets consisting of hundreds to thousands of individual observations. 
In particular, the Atacama Large Millimeter/submillimeter Array (ALMA) is a pioneering observatory that provides high-resolution observations at mm and sub-mm wavelengths. 
Transforming the large volume of raw visibility (uv) data from ALMA into scientifically meaningful products is a significant task. This complex process includes several steps: calibration, imaging, and deconvolution.

ALMA has made remarkable progress in automating and ensuring high-quality calibration of interferometric data. 
These advances have been driven by the diligent efforts of the observatory staff and the success of ALMA's sophisticated pipeline \citep{hunterALMAInterferometricPipeline2023}, based on the Common Astronomy Software Applications (CASA) software \citep{mcmullinCASAArchitectureApplications2007}. 
As a result, ALMA provides its users with meticulously calibrated visibilities, laying the groundwork for further processing and scientific analysis. 
Despite the effective functionality of the pipeline, certain complications can arise during the calibration process, requiring manual intervention.
These complications can include unexpected radio frequency interference, atmospheric anomalies, or misbehaved antennas, all of which affect data quality.
In such cases, the expertise of astronomers becomes crucial to correct or flag these problems, ensuring the integrity and reliability of the final scientific products.

Over the years, several large programs (LPs) have exploited the capabilities of ALMA to study the cosmic evolution of gas and stars.
In particular, the ASPECS survey in the Hubble Ultra Deep Field (HUDF; \citealt{walterALMASpectroscopicSurvey2016, decarliALMASPECTROSCOPICSURVEY2016}) aimed to detect CO and [CI] in galaxies without preselection at $z=1-3$ using Bands 3 and 6. ASPECS covered an area of 4.6 arcmin$^2$ in Band 3 and 2.9 arcmin$^2$ in Band 6. 
PHANGS–ALMA \citep{leroyPHANGSALMAArcsecondCO2021} is a survey of CO(2–1) emission from 90 nearby galaxies with a typical angular resolution of $\sim1.5''$ and a total survey area of 1050 arcmin$^2$.
At higher redshifts, the REBELS survey 
\citep{bouwensReionizationEraBright2022} targeted 40 UV-bright galaxies at $z>6.5$, covering an area of 7 deg$^2$, and aimed to detect the [CII]$-$158$\mu$m and [OIII]$-$88$\mu$m lines as well as dust-continuum emission.
The ALPINE survey \citep{lefevreALPINEALMACIISurvey2020} was designed to study 118 star-forming galaxies (SFGs) at $4 < z < 6$. This survey targeted the [CII] line and continuum emission, covering an area of 25 arcmin$^2$. 
The CRISTAL survey \citep{solimanoALMACRISTALSurveyDiscovery2024} selected 25 SFGs with available HST imaging and high stellar masses from the ALPINE sample ($\log$ M$_{\text{star}}$/M$_{\odot}$ $\geq 9.5$). CRISTAL targets [CII] with a resolution of $\sim 0.2''$, which is higher than its parent sample's resolution of $\sim1''$. 
These surveys have played a key role in constraining the evolution of the molecular gas in galaxies across cosmic epochs. Using different molecular gas tracers shows that the evolution of the molecular gas mass density in the Universe aligns with the cosmic star formation history, providing insights into the process of gas accretion onto galaxies \citep{walterEvolutionBaryonsAssociated2020}.
In addition to the ALMA LPs, the ALMA Calibrator source catalogue has also been used by \cite{audibertCOALMARadioSource2022} on a representative sample of the NVSS (when flux limited to 0.4 mJy) to study the CO luminosity function up to redshift $z \sim 2.5$ and to assess the role of radioactivity in galaxy evolution. They found that most radio galaxies are more depleted and evolved than the typical simulated halo galaxy.

Other facilities have also been used for surveys of cold gas tracers.
The Plateau de Bure High$-z$ Blue Sequence Survey 2 (PHIBSS2)   \citep{guilloteauIRAMInterferometerPlateau1992}, is a large observational campaign conducted with the Plateau de Bure Interferometer (PdBI), now part of the Northern Extended Millimeter Array (NOEMA) Observatory. 
With a frequency range spanning from 80 GHz to 350 GHz, \citet{lenkicPlateauBureHighz2020} searched for additional background emission lines in the fields of the original PHIBBSS survey. 
They explored the CO($2-1$), CO($3-2$), and CO($6-5$) emission lines in 110 main-sequence galaxies, covering a total area of $\sim130$ arcmin$^2$.
This survey has been used to derive the molecular gas mass density evolution by converting high$-$J CO luminosity functions to CO($1-0$).
Additionally, the COLDz survey looked directly for the CO($1-0$) emission at $z= 2-3$ and CO($2-1$) at $z = 5-7$ using more than 320 hrs of VLA time \citep{pavesiCOLuminosityDensity2018, riechersCOLDzShapeCO2019, riechersCOLDzHighSpace2020}, covering a $\sim 60$ arcmin$^2$ area. 
However, the large uncertainties of these measurements reflect our limited understanding of the molecular gas content of galaxies across cosmic time.

Our current measurements of the molecular gas mass density ($\Omega_{\text{H}_2}$) reach redshift $z\sim 7$ and show that the density increases from early times, peaks at $z\sim 1-3$, and decreases to the present day \citep[e.g.][]{tacconiHighMolecularGas2010, walterEvolutionBaryonsAssociated2020, hamanowiczALMACALVIIIPilot2022, aravenaALMAReionizationEra2023}.
However, one of the main challenges for the molecular gas surveys is the effect of cosmic variance, which introduces uncertainty on how well the sampled volume represents the Universe \citep{decarliALMASpectroscopicSurvey2020, poppingALMASpectroscopicSurvey2020, boogaardNOEMAMolecularLine2023}. 
Cosmic variance, caused by natural fluctuations in the Universe's large-scale structure, results in variations in the number density and distribution of cosmic objects across the sky \citep{driverQuantifyingCosmicVariance2010}. 
Consequently, surveys covering limited sky areas may inadvertently sample regions that are unusually rich or devoid of galaxies, leading to biased estimates of cosmic properties such as the number density of galaxies, clustering, and luminosity function. 
Small sample sizes exacerbate this issue by introducing statistical uncertainties, which increases the probability of sampling regions with atypical characteristics.
Several studies have investigated the effects of field-to-field variance on observables such as the luminosity function and the number density of galaxies \citep{keenanBiasesCosmicVariance2020, lenkicPlateauBureHighz2020, gkogkouCONCERTOSimulatingCO2022, boogaardNOEMAMolecularLine2023}. Generally, there is a consensus on the significance of measuring the cosmic variance effect by comparing model predictions with observations and estimates from various sky regions \citep{poppingALMASpectroscopicSurvey2019}. 
This is particularly crucial in deriving the molecular gas mass density of the Universe.

This paper presents a new survey based on ALMA calibrator data accumulated up to May 2022, which we denominated as ALMACAL$-22$. 
The original ALMACAL project started in 2016 \citep{oteoALMACALExploitingALMA2016} and has produced several scientific outcomes dedicated to studying molecular gas, dusty star-forming galaxies (DSFGs), absorption lines along the line of sight of quasars, and active galactic nucleus (AGN) physics.
ALMACAL$-22$ was built on the experience of previous pilot ALMACAL surveys, but it has now been expanded to include longer integration times. As part of the strategy to exploit this large dataset, we present the details of processing and imaging data, along with different tests, to ensure the best quality of the sample selection.
Here, we provide the characteristics and properties of this new dataset. We compare the strengths of ALMACAL$-22$ to previous surveys. 
This new survey covers 1047 fields across the southern sky, intending to alleviate the limitations introduced by Poisson errors due to limited statistics and cosmic variance. 
We review various studies conducted since ALMA Cycle 1 that have used extragalactic calibration data to explore interesting scientific cases \citep[e.g.][]{oteoALMACALIIExtreme2017a, klitschALMACALAbsorptionselectedGalaxies2019, hamanowiczALMACALVIIIPilot2022, chenALMACALIXMultiband2022}. We discuss how this new release, ALMACAL$-22$, can expand our understanding of molecular gas evolution, properties of DSFGs, extragalactic absorption lines, and AGN physics.

This paper is organised as follows. 
In Sect. \ref{sec:almacal} we describe the ALMACAL$-22$ survey, the calibration process (\S\ref{sec:calibratio}), the selection of the pruned sample (Sect. \ref{sec:clean_sample_selection}), and the concatenation and imaging (Sect. \ref{sec:imaging}). 
In Sect. \ref{sec:properties} we explore the following properties: spatial distribution (Sect. \ref{sec:spatial_distribution}), spatial resolution (Sect. \ref{sec:spatial_resolution}), integration time (Sect. \ref{sec:integration_time}), and calibrator redshifts (Sect. \ref{sec:redshift}). 
Section \ref{sec:discussion} compares the strengths of ALMACAL$-22$ with the previous surveys and discusses the scientific areas where ALMACAL$-22$ will significantly contribute. 
In Sect. \ref{sec:conclusions} we summarise our key conclusions. 
Throughout this paper, we use $H_0 = 70$ kms$^{-1}$Mpc$^{-1}$, $\Omega_{\text{M}} =0.3$, and $\Omega_{\Lambda} = 0.7$.

\section{ALMA calibrator data}
\label{sec:almacal}
The ALMACAL$-22$ survey comprises archival data compiled from the calibration data used in every science project carried out by ALMA \citep{zwaanALMACALSurveyingUniverse2022}.
Each PI-led scientific project involves several observations of a calibrator source that is close to the science field. 
Most calibrators are bright sub-mm point sources classified as blazars \citep{bonatoALMACALIVCatalogue2018a} --- AGN galaxies with a jet pointing towards the line of sight \citep{urryUnifiedSchemesRadioLoud1995}. Blazars can be divided into two sub-classes: BL Lac objects, which are identified as radio galaxies, and flat-spectrum radio quasars (FSRQs), which are identified as quasars \cite{padovaniTwoMainClasses2017}.
When targeting a source with unknown structure and flux, calibration observations are crucial in interferometric astronomy. We used calibrator sources with well-known shapes and flux densities at (sub-)mm and radio wavelengths to adjust the bandpass response, set the flux density scale, and calibrate the amplitude and phase. This ensures we can correct for instrumental and atmospheric corruptions, providing accurate measurements of the target source.
These calibration scans have exposure times and setups matching the project's PI requests. Repeated use of the calibrators for ALMA science observations effectively creates deep observations that often cover a significant fraction of the sub-mm spectrum. 
The most popular calibrators have data from multiple observations of different ALMA bands. 
Multiple observations of ALMA calibrators create a high-sensitivity dataset that covers a significant sky area ($> 1000$ arcmin$^2$).

\begin{figure}
        \includegraphics[width=1.\columnwidth]{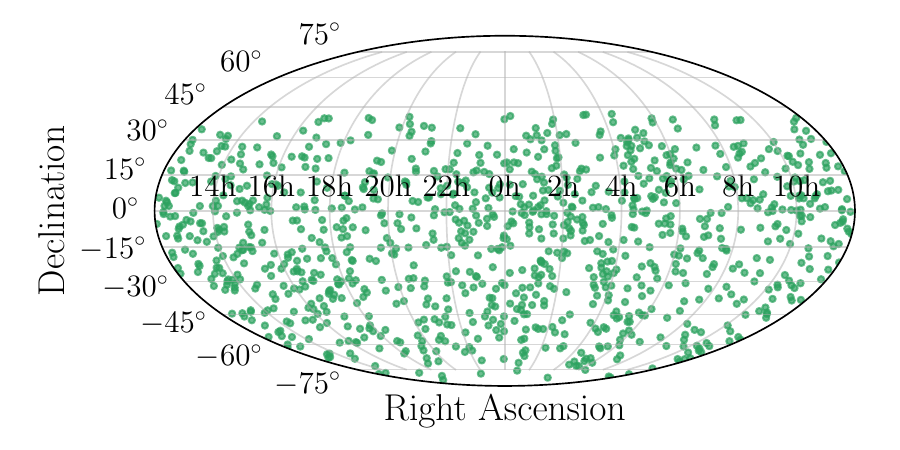}
     \includegraphics[width=1.\columnwidth]{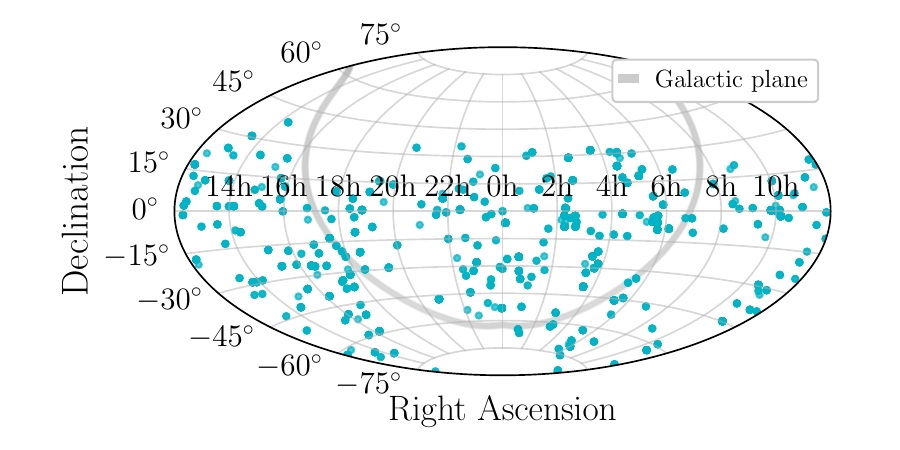}
    \caption{Spatial distribution of the calibrator fields for the full ALMACAL$-22$ (top) and pruned samples (bottom). The distribution of the full sample has no preferred direction, so the position of the calibrators in the sky can be considered homogeneous and only affected by the fact that ALMA observes the sky below $\delta<40^{\circ}$. 
    In the pruned sample, two regions exhibit sparse data, possibly attributed to interference, the limited availability of deep fields, and observational difficulties stemming from their proximity to the galactic plane. Further details are provided in Sect. \ref{sec:spatial_distribution}.}
    \label{fig:spatial_distribution}
\end{figure}

This paper presents 1047 calibration field data points accumulated from Cycle 1 (July 2012 - May 2022) from Band 3 through Band 10, resulting in a dataset of more than 30 Terabytes. 
These calibrator data are accessible to any user right after the main science dataset has passed quality assurance. In this section we explain the calibration process and subtraction of the calibrator at the centre of the pointing as well as the selection of the so-called pruned sample. Figure \ref{fig:spatial_distribution} shows the ALMACAL$-22$ field distribution on the sky for the full sample (top) and the pruned sample (bottom).

\subsection{Calibration} \label{sec:calibratio}

A dedicated ALMACAL pipeline was created to automate the processing of all delivered datasets. 
The complete pipeline description is available in \citet{oteoALMACALExploitingALMA2016}, and the following is a brief overview of its essential components.
This pipeline uses the \texttt{scriptForPI.py} script included in every delivered ALMA dataset, which generates fully calibrated data. While ALMA users use this script to create a final dataset, we used it to distinguish the calibrated calibrator data from the science observations.
Next, we applied self-calibration to correct the short-time variability of the phase and amplitude during integration, improving the final image's dynamic range. This process enabled us to create a model and remove the bright calibrator from the visibility data, resulting in the equivalent of blank sky or deep field observations.

During the execution of the ALMA calibration scripts, some bandpass calibrators needed to be consistently calibrated. Unlike with phase calibrators, calibration tables are not always applied to the bandpass calibrators. 
Most of the time, the bandpass calibrator is also the flux calibrator, so their solutions cannot be applied to themselves. 
To rectify this, the accurate flux density scale of the bandpass calibrators has to be recovered from the flux calibration table containing standardised values for each calibrator. 
Applying bandpass calibration tables to bandpass calibrators could, in principle, remove emission and absorption features from cubes that are constructed from these bandpass calibration observations. 
This could affect emission and absorption features at the phase centre. 
It should be noted, however, that the bandpass solution is spectrally averaged so that narrow spectral features are not calibrated out. 
Also, our ability to detect faint emission lines throughout the cubes is unaffected. 
Because emission lines on top of the quasar continuum are weak compared to the quasar continuum, they do not significantly affect the bandpass solution and do not affect the calibration quality of the data.

The calibration process extends to creating "pseudo-continuum" measurement sets, in which all channels within each spectral window are averaged. 
The averaging also boosts the signal-to-noise ratio (S/N) for the self-calibration.
These files facilitate the implementation of two steps of self-calibration, initially focusing solely on phase and subsequently incorporating both amplitude and phase calibration. 
The solution interval for the calibration solution is set equal to the integration time. 
During the intervals of the self-calibration steps, a point source model is applied and fitted to the {uv} data.
This approach offers the advantage of subtracting the point-source model independently for each observation, mitigating the impact of any flux variability in the calibration source. 
The final step involves deriving the calibrated visibilities and subtracting the continuum of the calibrator source to build line emission data cubes, thereby assembling the ALMACAL$-22$ dataset.

\begin{figure}
    \includegraphics[width=0.85\columnwidth]{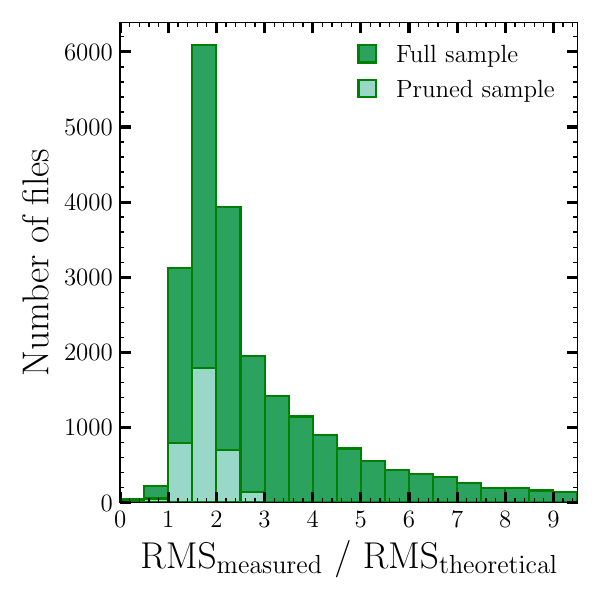}
    \caption{Distribution of the RMS of measured flux values in maps over the theoretical ones for each file in the full ALMACAL$-22$ and pruned samples. The pruned sample has a median of 1.8, reaching the imposed limit of 3, whereas the full sample's median is 2.4. The full sample also includes large values in the hundreds, although we truncate the figure at a ratio of 10. It is necessary and important to remove problematic files that may compromise the final product once these files are combined.}
    \label{fig:rms_dist}
\end{figure}

\subsection{Pruned sample selection}
\label{sec:clean_sample_selection}
While the central bright calibrator enables self-calibration, it introduces challenges to the scientific analysis. 
Most of the calibrators in our sample are blazars, but a few present extended structures that cannot be modelled as point sources. 
After subtracting the calibrator's continuum at the centre of the pointing as a point source, these structures may remain as residuals since they are not taken into account during the cleaning process.
These residuals affect the quality of the calibration process, resulting in strong 
continuum artefacts, interferometric patterns, or extra features that produce high noise values.
To overcome this problem, we implemented a sequence of pruning steps to prevent any negative impact on the final combined images.

We analysed key properties, such as the integration time, frequency coverage, bandwidth of each spectral window, spatial resolution, and baseline distribution. We aim to obtain a coherently combined homogeneous sample that will provide broad statistics of the Universe, minimise cosmic variance, and cover the maximum possible volume.

The pruned sample includes only observations made with the 12-metre array.
To create the deep cubes, we summed the on-source integration time of observations that covered the same frequency range. 
We chose bins of 1 GHz to select all the files spanning the same frequency across the extent of the ALMA band's frequency coverage.  
To minimise contamination from artefacts, we used a multi-step approach. 
First, we created the continuum image for each measurement set using \texttt{CASA} and estimated the root-mean-square (RMS) noise. We chose to use the continuum image instead of the data cube because imaging the continuum is much faster and provides a good approximation of the quality of the files.
We excluded files with an RMS value higher than $10^{-5}$ Jy, which introduced inconsistent noise patterns throughout the cube. 
Secondly, we estimated the theoretical sensitivity ($\sigma_{\text{theoretical}}$) associated with each observation by using the \texttt{apparentsens} function of the \texttt{CASA} image \texttt{toolkit}. This function calculates the point source sensitivity in imaged cubes, accounting for image weights and visibility weights,
also used in the ALMA Interferometric Pipeline \citep[see][]{hunterALMAInterferometricPipeline2023}.
We insist that the measured RMS of each continuum imaged file compared with the theoretical sensitivity (ALMA pipeline) should not be greater than a factor of 3. 
We inspected a few files that were removed by the RMS criteria and found that most of them had strong interferometric patterns or errors in calibration. 
Figure ~\ref{fig:rms_dist} shows the distribution ratio of the measured RMS and the theoretical RMS. 
The pruned sample extends by up to a factor of 3 with a median value of 1.8, while the full sample has a median value of 2.4. 
For comparison, the full sample reaches RMS ratios of hundreds. These high-RMS observations could introduce significant noise artefacts in the final data product.

Despite our efforts with the sample selection, we still found files with strong interferometric patterns, such as image artefacts, baseline stripes, blurring, or distortions in the image. 
In some cases, these patterns cannot be easily identified from the RMS of the continuum.
Several errors can arise when one performs self-calibration using the pipeline, such as poor phase and amplitude calibration. 
These artefacts are usually caused by baseline errors due to uncertainties in the antenna position when measuring the differences in signal phase and amplitude between pairs of antennas. 
Artefacts can also appear during imaging, resulting in sidelobes, noise bias, and spatial distortion. 
The errors in interferometric data are primarily identified in the {uv} plane during calibration and imaging processes.
Considering the large amount of data that the ALMACAL$-22$ survey contains, it is not possible to visualise all the observations in the {uv} plane to flag the possible misbehaved antennas. 
Thanks to our large amounts of data, we can afford to discard corrupted data by checking the continuum map of each observation to determine if it presents strong patterns. 
We visually identified the presence of stripes, blurring, artificial elongation or shifting of sources, and symmetric or asymmetric patterns, then discarded observations exhibiting these issues to ensure data quality.
After removing $\sim 15 \%$ of the files, we combined the data with the same frequency coverage, selecting observations that contribute to a total integration time of at least 10 minutes.

\begin{figure*}
        \includegraphics[width=0.65\columnwidth]{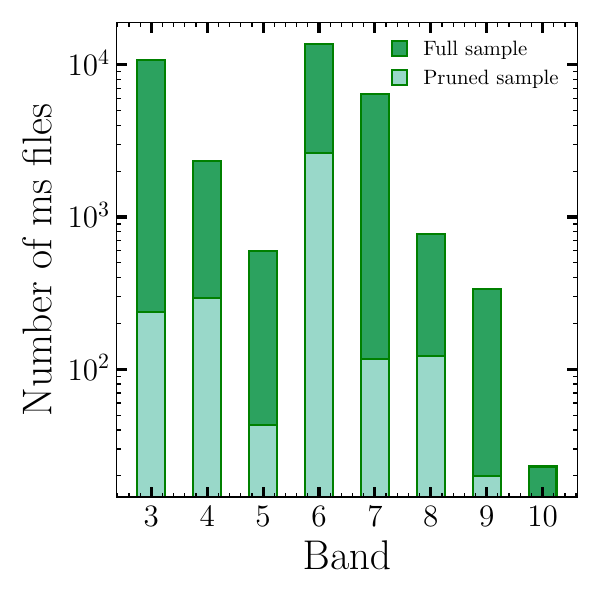}
    \includegraphics[width=0.65\columnwidth]{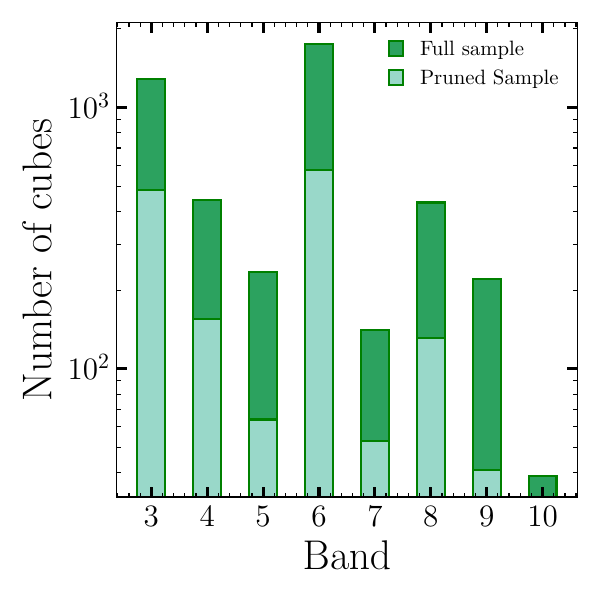}
    \includegraphics[width=0.65\columnwidth]{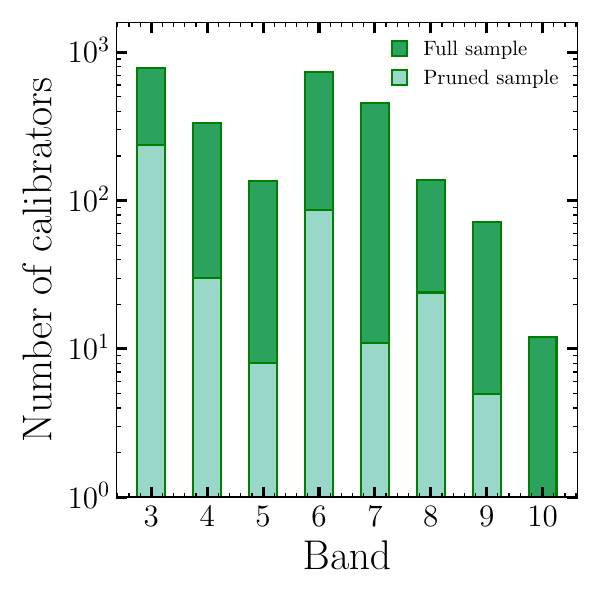}
    \caption{Distribution of the number of ms files (left), cubes (middle), and calibrators (right) for the full ALMACAL$-22$ and pruned samples across ALMA bands. The full sample contains 34909 ms files, 4547 cubes, and 1047 calibrators, while the pruned sample consists of 6494 ms files, 1508 cubes, and 401 calibrators. Band 6 comprises most files and cubes for the full ALMACAL$-22$ and pruned samples. Band 3 has the largest variety of calibration fields for both samples, followed by Bands 6 and 4. It should be noted that no data are available in Band 10 for the pruned sample as the quality assurance discarded all the data that contributed sufficient integration time (10 minutes).}
    \label{fig:subsample_selection}
\end{figure*}

We selected high-quality data to construct the survey. 
We started from a total of 34909 measurement sets (ms), that is, raw visibility files, of which 25594 add up to at least 10 minutes of integration time. 
Among these, 15270 files have acceptable RMS levels. After removing visual artefacts and reapplying the integration time criterion, we used 6494 ms files to build the pruned sample.
Figure~\ref{fig:subsample_selection} displays the distribution of the number of files in each band for both the full ALMACAL$-22$ and pruned samples in the left panel. 
Bands 3, 6, and 7 dominate the full sample, while Bands 3, 4, and 6 dominate the pruned sample. 
There are no Band 10 data available in the pruned sample.
The middle panel shows the distribution of data cubes in both samples, where Bands 3, 4, 6, and 7 dominate (details on how cubes are constructed are discussed in Sect. \ref{sec:imaging}).
The right panel shows the distribution of calibrators, with 1047 in the full sample and 401 in the pruned sample. For both samples, Band 3 has the most calibration fields, followed by 6 and 4.

\begin{table}
        \centering
        \caption{Number of data cubes and calibrators in the pruned sample}
        \label{tab:n_cubes_cal}
        \begin{tabular}{cccc} %
                \hline \hline
                Band & Frequency  & Nº data cubes & Nº calibrators \\
        & [GHz] & & \\
                \hline
                3 &$84 - 116$ & 485 & 237 \\
                4 &$125 - 163$ &155 & 75 \\
                5 & $163-211$ &64 & 31 \\
        6 & $211 - 275$ &579  & 201  \\
        7 & $275-373$ & 53 & 19  \\
        8 & $385-500$ &131 &  57 \\
        9 & $602 - 720$ & 41 &  15 \\
                \hline \hline
        \end{tabular}
\end{table}

\subsection{Building cubes}
\label{sec:imaging} 

Data cubes have three dimensions, two of which contain spatial information and one spectral information given by a frequency range.
To produce each data cube, we followed a sequence of steps to combine multiple observations. First, we concatenated all the {uv} observations into a single file to obtain a data cube.
We selected the ms files with sequential frequencies to be included in the cube.
Then we applied the \texttt{CASA} task \texttt{statwt} to the concatenated file to recalibrate the {uv} weighting of different observations based on the variance of data.
We estimated the beam size using the \texttt{CASA} task \texttt{getsynthesizedbeam} from the \texttt{analysisUtils} package. 
We sampled each beam with 3 pixels (using a pixel scale = synthesised beam / 3) and picked the image size to be $\sim1.8$ times that of the primary beam, selecting the number of pixels recommended by \texttt{CASA} to maximise the imaging efficiency.
We defined the channel width for all cubes to be 31.2 MHz, a value selected to strike a balance between spectral resolution, S/N, and manageable data volume.
Finally, we ran \texttt{tclean} in \texttt{CASA} for imaging using linear interpolation, natural weighting to optimise sensitivity, and 0.5 arcsec tapering.
We parallelised \texttt{CASA} using eight cores and eight threads, which reduces by a factor of 4 the execution time of \texttt{tclean} for some cubes. 
The duration of the imaging process depends on how many observations we combine and the cubes' frequency coverage, ranging from a few minutes to several hours per cube.

The pruned sample is composed of 1508 cubes from Band 3 to Band 9. Table \ref{tab:n_cubes_cal} details the number of cubes constructed in each band for the pruned sample and the number of calibrators covered. 
The distribution of the number of cubes is shown in the middle panel of Fig. \ref{fig:subsample_selection}. 
The full sample exhibits three times more cubes than the pruned sample. 
The number of cubes in the full sample refers to the number of data cubes that can be built applying the integration time criterion only, that is, without taking into account the RMS selection and visual inspection that was used to build the pruned sample (see Sect. \ref{sec:clean_sample_selection}). 

\section{ALMACAL$-22$ properties} \label{sec:properties}
This section explores the inherent properties of the latest data from the ALMACAL$-22$ survey in both the full and pruned samples.
We characterise the fundamental properties of this dataset to provide a clear understanding of the survey's scope and capabilities.
These include the spatial distribution (Sect. \ref{sec:spatial_distribution}), spatial resolution (Sect. \ref{sec:spatial_resolution}), integration time (Sect. \ref{sec:integration_time}), and redshift of the calibrator sources (Sect. \ref{sec:redshift}). 

\subsection{Spatial distribution} \label{sec:spatial_distribution}

The ALMACAL$-22$ survey comprises calibrators randomly distributed across the southern sky, resulting in a widely dispersed area covering more than 1100 arcmin$^2$. 
The data collection strategy effectively captures diverse regions with a random distribution. 
This distribution is advantageous for serendipitous line and continuum detections, providing robustness against the effects of cosmic variance, a challenge often encountered in deep field surveys. 
The top panel of Fig. \ref{fig:spatial_distribution} shows the spatial distribution for the full ALMACAL$-22$ sample and bottom panel the pruned sample.
In the pruned sample, two areas have relatively sparse data. These regions are near the projection of the galactic plane, likely resulting in fewer observations due to potential interference and the relative paucity of cosmological deep fields in those regions, or challenges associated with observing in those directions. 
The primary factor influencing the selection of the pruned sample is the minimum integration time, which determines how long observations need to be made to qualify for inclusion in the dataset.

\subsection{Spatial resolution} \label{sec:spatial_resolution}

The spatial resolution of the ALMACAL$-22$ dataset varies depending on the observing frequency and array configuration chosen by PIs based on their scientific goals.
Higher-frequency bands generally offer a finer spatial resolution, enabling the detection of more intricate structures and providing valuable insights into the morphology of the observed sources. 
Conversely, lower-frequency bands offer advantages regarding sensitivity and wider coverage but with a potential sacrifice of some spatial resolution.

\begin{figure*}
\includegraphics[width=0.5\columnwidth]{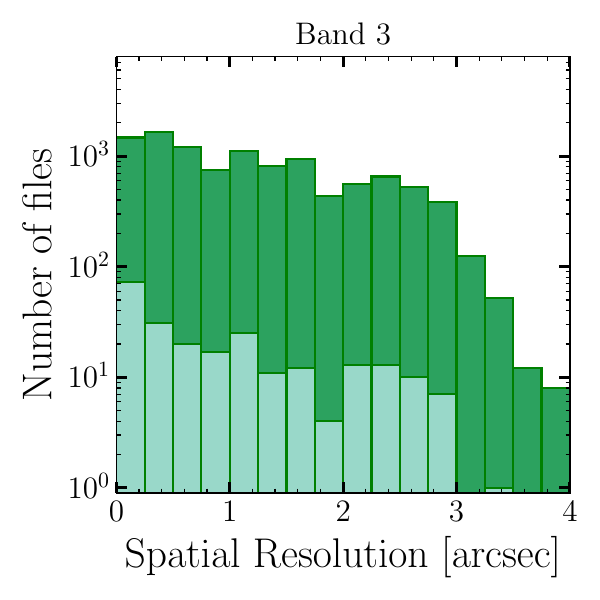}
\includegraphics[width=0.5\columnwidth]{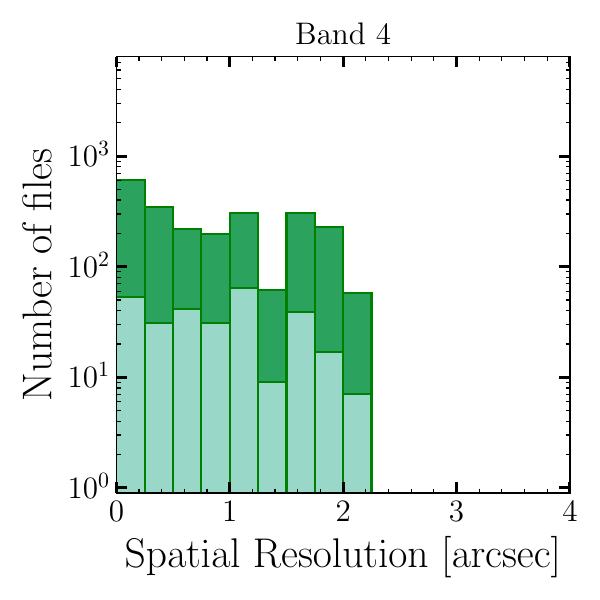}
\includegraphics[width=0.5\columnwidth]{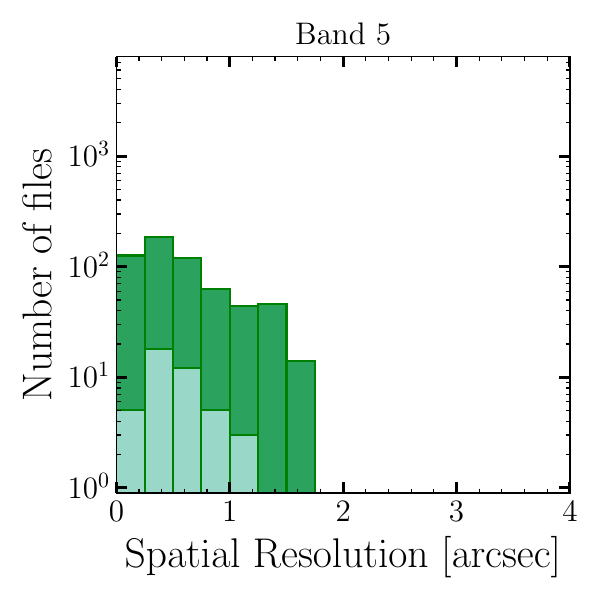}
\includegraphics[width=0.5\columnwidth]{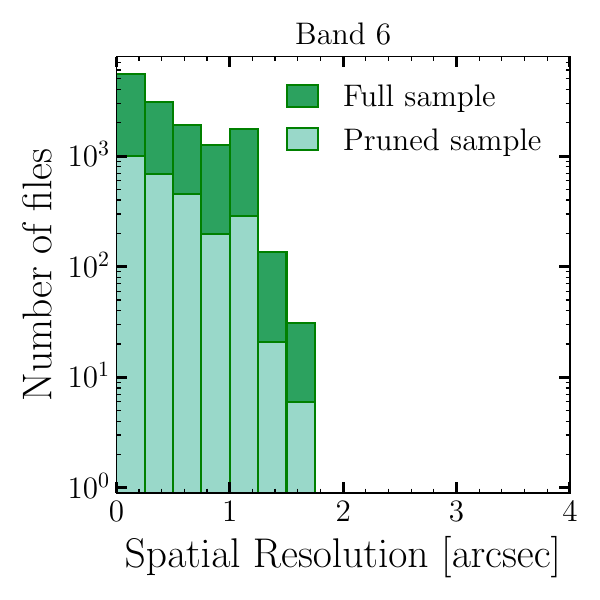}
\includegraphics[width=0.5\columnwidth]{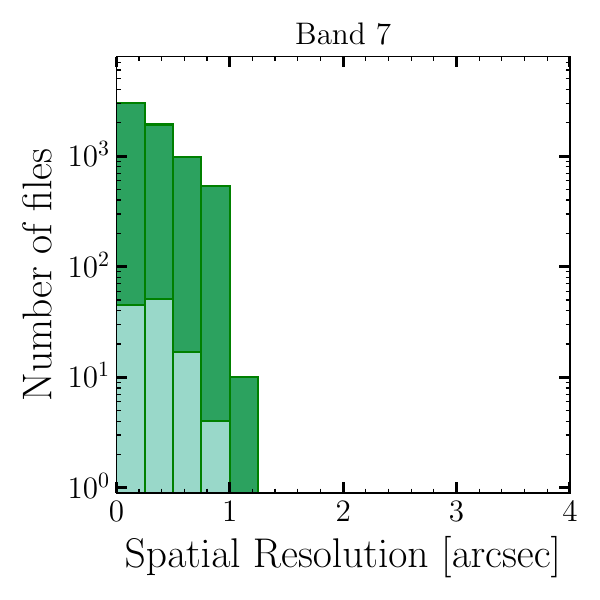}
\includegraphics[width=0.5\columnwidth]{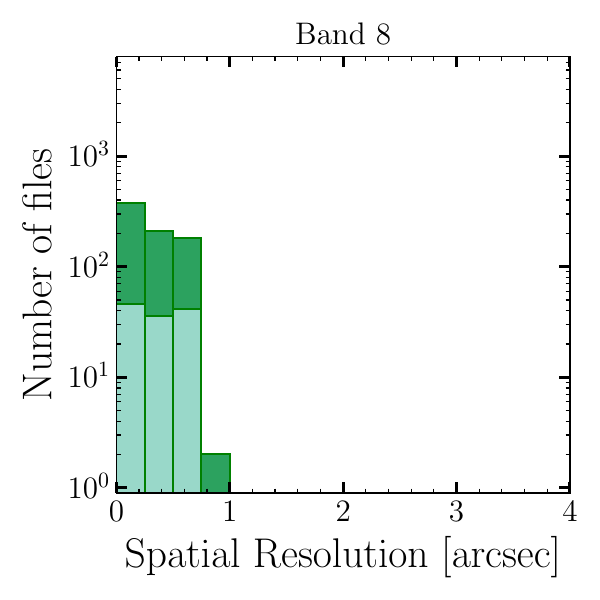}
\includegraphics[width=0.5\columnwidth]{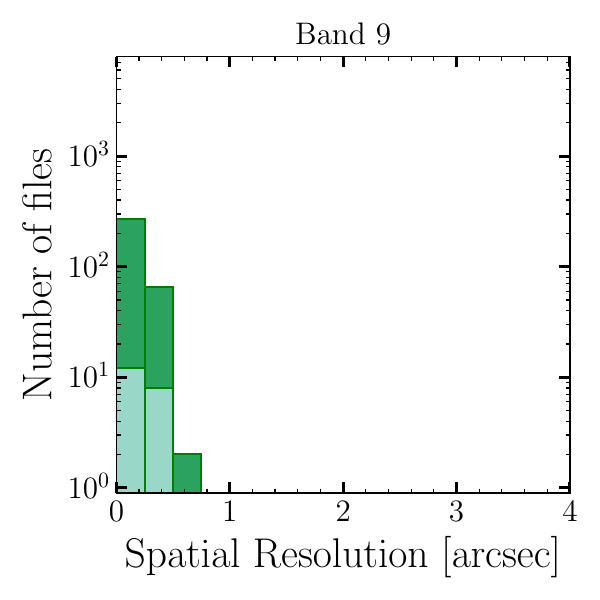}
\includegraphics[width=0.5\columnwidth]{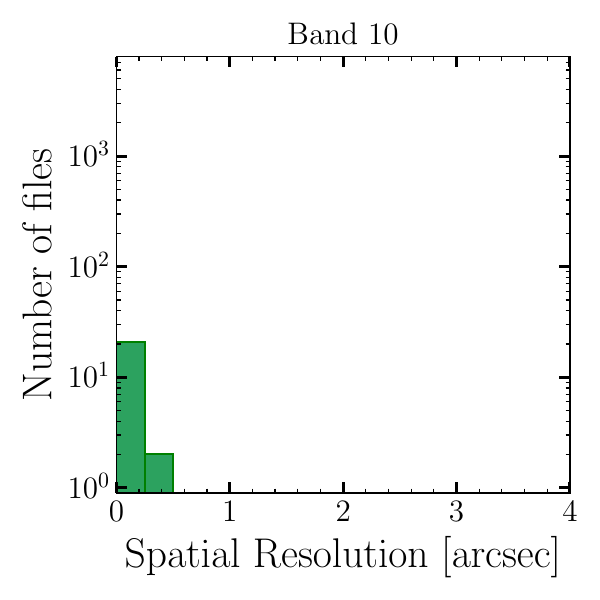}
    \caption{Distribution of the spatial resolution for the full ALMACAL$-22$ and pruned samples across ALMA bands. The mean values of the full ALMACAL$-22$ sample decrease consistently towards higher bands, from a median value of 1.15 arcsec in Band 3 to 0.10 arcsec in Band 10. In the pruned sample, Band 4 presents a median value of 0.87 arcsec, followed by Band 3 with 0.59 arcsec, and decreases in higher-frequency bands, reaching 0.23 arcsec in Band 9.}
    \label{fig:spatial_resolution}
\end{figure*}

Figure \ref{fig:spatial_resolution} shows how each band's spatial resolution is distributed.
In the ALMACAL$-22$ full sample, the median value in Band 3 is 1.14 arcsec. This consistently decreases for higher-frequency bands, reaching a median spatial resolution of 0.13 arcsec in Band 10. In the case of the pruned sample, Band 4 shows the maximum value in the median spatial resolution of 0.87 arcsec. This value is lower in Band 3, then decreases to a mean value of 0.23 arcsec in Band 9.

\subsection{Integration time} \label{sec:integration_time}
The integration time reflects the total on-source observation time obtained for the ALMA calibration data.
The dataset's construction requires combining two or more observations if their frequencies overlap by at least 1 GHz.
We added up the integration time in a 1 GHz moving window, selecting all the files spanning the same frequency. We imposed a lower limit of 10 minutes for the pruned sample across the whole frequency coverage.
Variations in integration time in the cubes arise from the concatenation of short calibration pointings, each lasting several minutes with overlapping spectral coverage. 
From the varying depths reached, different sensitivity values can be achieved within a cube, but every frequency meets the minimum time criterion.
The cube's frequency range within a cube is chosen to contain sequential frequencies. 
A calibration field observed several times in the same frequency band may have different frequency ranges observed,
creating more than one cube for the same field and band, covering a distinctive frequency extent.

\begin{figure}
        \includegraphics[width=0.85\columnwidth]{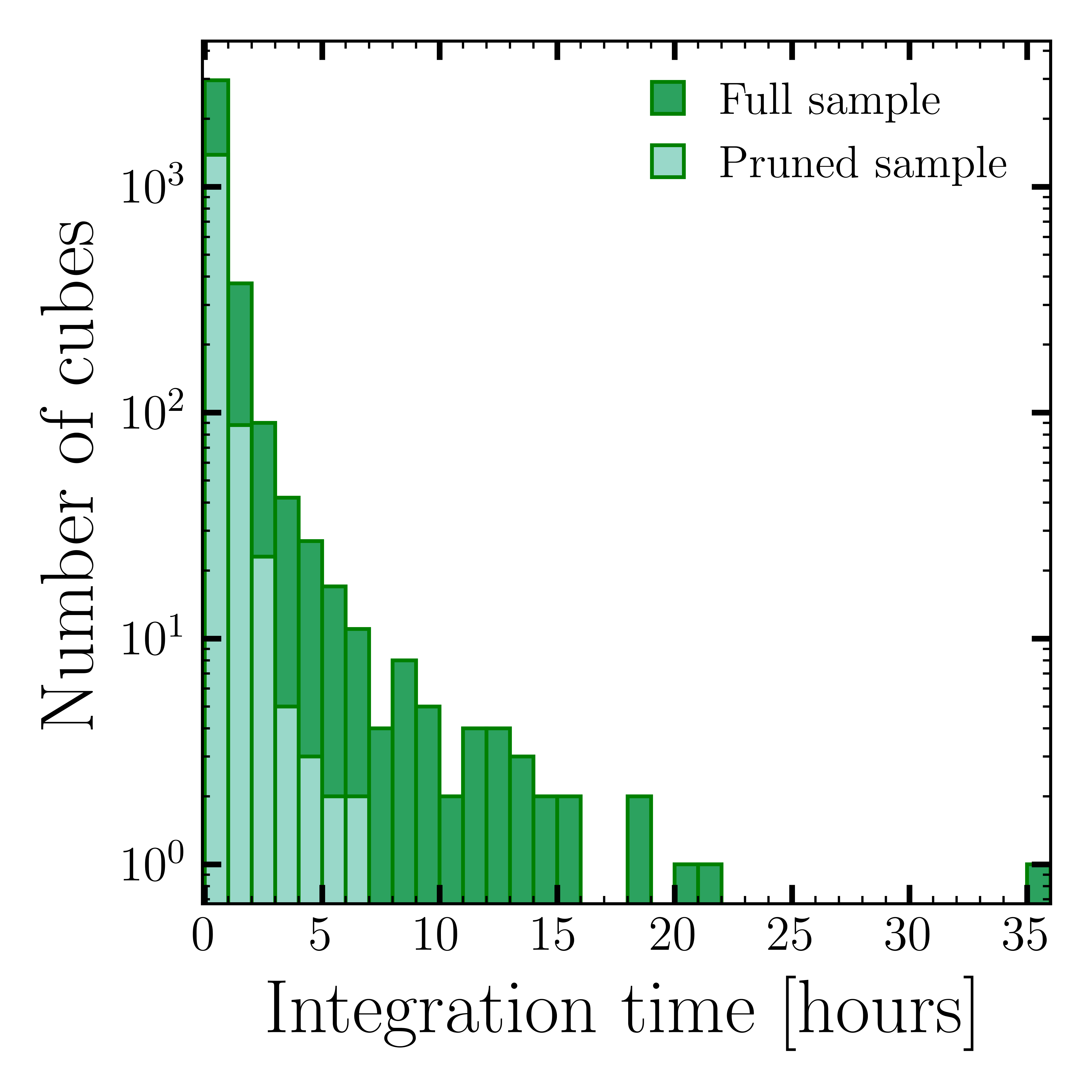}
    \caption{Distribution of the maximum integration time reached in each data cube for the full ALMACAL$-22$ sample and the pruned sample. The maximum and median values reached in each cube for each band are listed in Table \ref{tab:int_time}. The full sample has a mean value of 0.78 hours, while that of the pruned sample is 0.48 hours. The pruned sample achieves up to 7 hours of integration time, maintaining high sensitivity levels.}
    \label{fig:int_time}
\end{figure}

Figure \ref{fig:int_time} shows the distribution of total integration times for the full ALMACAL$-22$ dataset and the pruned sample. 
The median and maximum values of the total integration times for the full ALMACAL$-22$ dataset and the pruned sample are shown in Table \ref{tab:int_time}. 
The full sample comprises cubes that achieve more than 10 hours of integration time, while the pruned sample reaches an integration time of up to 7 hours. 
The median value in the full sample is 0.78 hours; in the pruned sample, it is 0.48 hours. 
The pruning process has only a slight impact on the median value, suggesting a small effect on the sensitivity. The mean sensitivity value reached in the pruned sample is $\sim0.78$ mJy/beam (for further discussion see Sect. \ref{sec:area_sensitivity}).

\begin{table*}[tb]
        \centering
        \caption{Main properties of the full ALMACAL$-22$ sample and the pruned sample.  }
    
        \label{tab:int_time}
        \begin{tabular}{ccccc|cccc} 
                \hline \hline & & & & & & & & \\
                 & &  Full sample   & &  & & Pruned sample & &  
        \\ & & & & & &  & & \\
                \hline 
         Band & Median $t_{\text{int}}$ & Max $t_{\text{int}}$ & Median Res. & Area & Mean $t_{\text{int}}$ & Max $t_{\text{int}}$ & Median Res. & Area \\
         & [min] & [h] & [arsec] & [arcmin$^2$] & [min] & [h] & [arcsec] & [arcmin$^2$] \\  
         \hline 
                3 & 19.7 & 21.2 & 1.15 & 1301.7  &
        19.0 & 6.1 & 0.59 & 693.1 \\
                4 & 19.4 & 5.9 & 0.75 & 209.2 &   
        14.8 & 2.7 & 0.87 & 250.8  \\
                5 & 28.1 & 2.3 & 0.49 & 54.7 &
        13.5 & 1.0 & 0.48  & 63.1 \\
        6 & 23.0  & 35.3 & 0.38 & 231.2 & 
        19.4 & 6.9 & 0.39 & 103.6 \\
        7 &  22.2 & 18.3 & 0.28 & 71.2 & 
        14.6 & 2.0 & 0.29  & 16.2 \\
        8 & 21.8 &  5.8 & 0.25 & 15.0 &   
        19.2 & 5.2 & 0.34  & 26.1 \\
        9 & 23.4 &  6.4 & 0.15 &  3.9 &   
        20.2 & 0.78  & 0.23  & 3.2 \\
        10 &  23.8  &  0.5  &  0.10   & 
        0.34 & ... & ...  & ... & ... \\ 
                \hline \hline
        \end{tabular}
 \tablefoot{Columns: (1) ALMA band, (2) and (6) median integration time in the cubes of the full sample and the pruned sample, (3) and (7) maximum integration time reached in the cubes of the full sample and the pruned sample, (4) and (8) median spatial resolution of the full sample and the pruned sample, (5) and (9) total covered area by all the cubes in the full sample and the pruned sample.}
\end{table*}

\subsection{Redshift}
\label{sec:redshift}

\begin{figure}
\includegraphics[width=0.85\columnwidth]{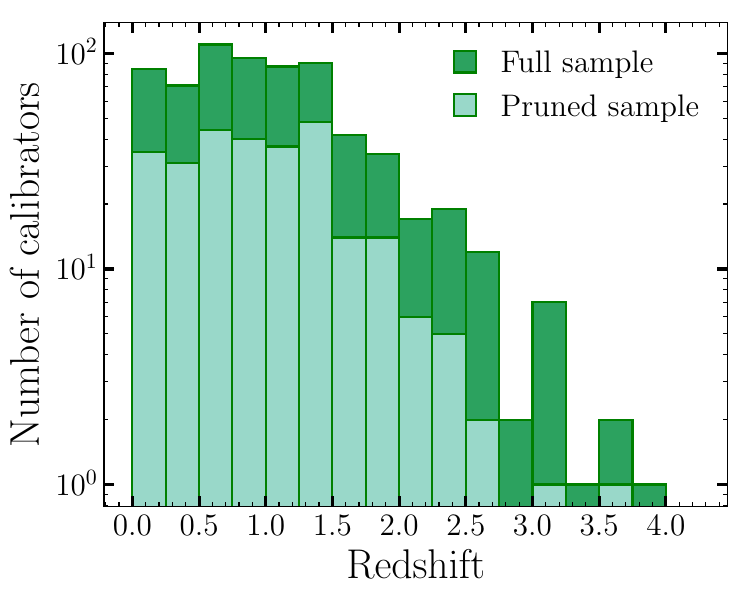}
\caption{Redshift distribution of calibrators in the full ALMACAL$-22$ sample and the pruned sample (Weng et al., in prep.). The median redshift is $\langle z \rangle = 0.749$ in the full sample, and $\langle z \rangle= 0.797$ in the pruned sample. The highest redshift calibrator is at $z \sim 3.8.$ }\label{fig:redshift_distribution}
\end{figure}
    
The ALMACAL Redshift Catalogue relies primarily on a redshift database introduced by \cite{bonatoALMACALIVCatalogue2018a} for a substantial portion of the calibrator sample.
This compilation was further enriched after cross-matching radio sources with optical sources from NED, SIMBAD \citep{wengerSIMBADAstronomicalDatabase2000}, and an optical catalogue of bright sources at gigahertz frequencies \citep{mahonyOpticalPropertiesHighfrequency2011}.
Team members individually checked each source, noting the redshift and provenance. This catalogue is also supplemented by VLT/X-Shooter observations of calibrator sources (ID 111.253L.001, PI: S. Weng and ID 0101.A-0528, PI: E. Mahony). 
Out of the initial 1047 sources, 675 have robust redshifts and spectra. For the pruned sample of this work, 390/635 ($61\%$) calibrators have confirmed spectroscopic redshifts. Further details will be presented in a forthcoming paper (Weng et al., in prep.). 

The redshift distribution for calibrator sources in the full ALMACAL$-22$ dataset and the pruned sample is shown in Fig. \ref{fig:redshift_distribution}. 
The full sample of calibrators has a median redshift of $z = 0.749$, with a maximum of $z=3.788$. 
In the pruned sample, the median redshift value is $z=0.797$, and the maximum is $z=3.591$. The sources excluded during pruning were likely more concentrated at lower redshifts. However, the redshift distribution in Fig. \ref{fig:redshift_distribution} shows that the higher redshift end is also less populated after pruning. Therefore, we do not expect any significant bias related to redshift in the pruned sample.

\section{Discussion} \label{sec:discussion}

This section delves into a comparative analysis of ALMACAL$-22$ and other surveys, emphasising key properties such as sensitivity and survey area. 
We provide an overview of the scientific goals the ALMACAL survey has achieved so far. 
We evaluate the unique characteristics of this extensive dataset compared to existing surveys, shedding light on its strengths and potential scientific contributions.

\subsection{Survey area and sensitivity} \label{sec:area_sensitivity}
We compared the sensitivity and total area covered by the ALMACAL$-22$ survey to those of previous large programs.
Due to the primary beam response, the sensitivity in each cube decreases as we move away from the phase centre. 
We considered an area 1.8 times the primary beam size for each field, which we calculated using the \texttt{primaryBeamArcsec} function from the \texttt{analysisUtils} package in \texttt{CASA}.
To estimate the sensitivity reached in each data cube, we used the \texttt{sensitivity} task from \texttt{analysisUtils}, which takes the integration time and central frequency as inputs.

ALMACAL$-22$'s coverage is distinguished by its larger footprint than any previous survey, covering over one thousand square arcminutes.
For comparison, the ALMA Spectroscopic Survey in the Hubble Ultra Deep Field (HUDF), ASPECS, \citep{decarliALMASPECTROSCOPICSURVEY2016, walterALMASpectroscopicSurvey2016, aravenaALMASpectroscopicSurvey2019, boogaardALMASpectroscopicSurvey2020}, covers 4.6 arcmin$^2$ area.
The COLDz survey \citep{pavesiCOLuminosityDensity2018, riechersCOLDzShapeCO2019, riechersCOLDzHighSpace2020} covers a $\sim 60$ arcmin$^2$ area.
The $6\times15$m IRAM Plateau de Bure High-z Blue-Sequence Survey 2 (PHIBSS2) covers a total area of $\sim130$ arcmin$^2$ \citep{tacconiPHIBSSUnifiedScaling2018, lenkicPlateauBureHighz2020}.

Regarding sensitivity, 
ALMACAL$-22$ data cubes achieve remarkable sensitivity levels in ALMA bands 3, 4, 5, 8, and 9, comparable to ASPECS and COLDz. 
Bands 6 and 9 offer a wider range of sensitivities, mainly due to the variety of total integration times obtained in each cube after combining data with overlapping frequencies. Additionally, \cite{audibertCOALMARadioSource2022} demonstrated that ALMA calibrator data can be used to estimate the molecular gas content of galaxies, reaching sensitivities around $0.482$ mJy, comparable to surveys such as COLDz and GOODS-ALMA.

\begin{figure}
    \includegraphics[width=1.2\columnwidth]{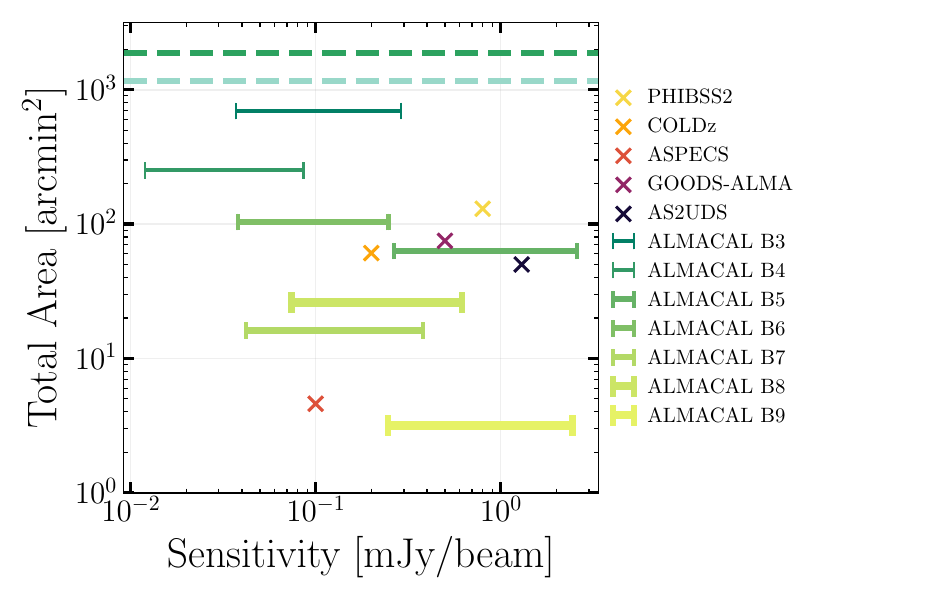} 
    \caption{Total ALMACAL survey area vs the median sensitivity range reached in each band for the pruned sample. For comparison, we plot the area and sensitivity values from the previous surveys, PHIBSS2 \citep{lenkicPlateauBureHighz2020}, COLDz \citep{riechersCOLDzShapeCO2019}, ASPECS \citep{decarliALMASPECTROSCOPICSURVEY2016} in ALMA Bands 3 and 6, GOODS-ALMA \citep{gomez-guijarroGOODSALMASourceCatalog2022} in ALMA Band 6, and AS2UDS \citep{stachALMASurveySCUBA22019} in ALMA Band 7. The dashed light and dark lines show the total area accumulated by summing up all the bands for the pruned sample and the full sample. The total survey area counts every calibration field once, considering the largest area.}
    \label{fig:area_sensitivity}
\end{figure}

Figure \ref{fig:area_sensitivity} compares the total survey area reached by each band in the pruned ALMACAL$-22$ sample as a function of the sensitivity.
For comparison, we also plotted the sky coverage and sensitivity values reached by other surveys \citep{lenkicCOExcitationHighz2023, riechersCOLDzShapeCO2019, decarliALMASpectroscopicSurvey2019, gomez-guijarroGOODSALMASourceCatalog2022, stachALMASurveySCUBA22019}.
Our pruned sample adds up to a total area of 1154 arcmin$^2$, shown as the light green dashed horizontal line, surpassing the total area covered by previous surveys.
As the dark dashed horizontal line shows, the full sample area over all bands reaches 1887 arcmin$^2$.
The total survey area for both the full and the pruned sample accounts for each calibration field once, prioritising the largest area.
Overall, ALMACAL$-22$'s combination of extensive area coverage and average sensitivity will complement findings from other large surveys, addressing the effect of cosmic variance.

Despite these strengths, there are challenges to scientific interpretation. 
For example, since ALMACAL consists of pointed observations of specific sources, it cannot be considered a truly blind survey, raising concerns about potential biases in cosmic overdensities. 
However, \citet{bonatoALMACALIVCatalogue2018a} classified most calibrators as blazars ($\sim90\%$).
This property mitigates clustering significance since the jets' brightness is due to their orientation effects toward the observer rather than their mass.
For instance, \cite{sushchProbingClusterEnvironments2015} found that clustering effects of a significant number of galaxies (more than five) situated near the line of sight of the blazar beam are absent in the local universe, but they may be possible at higher redshifts ($z > 2$). 
Moreover, the lack of correlation in redshifts between continuum- or line-detected galaxies and the calibrators further undermines the impact of having the calibrator source at the centre of each pointing \citep[see][]{hamanowiczALMACALVIIIPilot2022}. Additionally, \cite{chenALMACALXIOverdensities2023} explored an overdense region in ALMACAL similar to extreme proto-cluster cores and found the most likely explanation to be alignment effects. Furthermore, the smaller primary beams in higher ALMA bands probe smaller volumes at lower redshifts.

\subsection{Science projects} \label{sec:science_projects}

The ALMACAL survey covers many scientific studies, focusing on four main areas: (1) molecular gas evolution, (2) properties of DSFGs, (3) extragalactic absorption lines, and (4) AGN physics. We briefly summarise what has been done in these areas and the impact that ALMACAL will have through further analysis of this dataset.

(1) The evolution of molecular gas has been investigated with ALMACAL through the redshifted CO emission line. 
ALMA observations cover the frequency range of CO lines at different redshifts. 
\cite{klitschALMACALAbsorptionselectedGalaxies2019} searched for the CO line in absorption in gas-rich galaxies selected via quasar absorption lines.
They found multiple CO transitions, revealing that galaxies were associated with optically identified AGN activity. 
They reported different factors when using the CO spectral energy distributions (CO SLEDs) as a proxy to estimate the amount of molecular gas compared to the widely used galactic values. These findings indicate the galactic values might overestimate the molecular gas masses for some absorption-selected galaxies. This difference highlights the need to construct CO SLEDs in different systems rather than assuming the values measured for typical SFGs.
More recently, \citet{hamanowiczALMACALVIIIPilot2022} developed an ALMACAL-CO pilot program to detect CO emission lines blindly over 38 calibrator fields, selected to have the longest integration times ($> 40$ minutes). This pilot program aimed to probe the feasibility of using ALMA calibrator fields to look blindly for CO emitters. Eleven emission lines were detected, providing a consistent estimation of the evolution of molecular gas compared with previous surveys. ALMACAL's untargeted approach offers the advantage of being less sensitive to cosmic variance than previous deep surveys.

(2) Dusty star-forming galaxies represent more normal SFGs than conventional sub-mm galaxies; such faint systems are usually buried in the confusing noise of sub-mm wavelengths. Early on, \citet{oteoALMACALExploitingALMA2016, oteoALMACALIIExtreme2017a} exploited the high sensitivity levels achieved by ALMACAL by combining data from multiple visits to 69 ALMA calibrator fields in Bands 6 and 7. 
They found eight DSFGs and derived the number counts. 
They discovered systems so faint that even the deepest Herschel surveys would not have detected them.
\cite{klitschALMACALVIIFirst2020} reported the first number counts in ALMA  Band 8 over 81 calibrator fields, finding 21 DSFGs. 
Recently, \cite{chenALMACALIXMultiband2022} extended the number counts estimation using 1001 calibration fields in the ALMA Bands 3, 4, 5, 6, and 7, covering a wavelength range from 3mm to 870 $\mu$m. 
They report the detection of 186 DSFGs with flux densities comparable to existing large ALMA surveys, but less prone to cosmic variance. Establishing the space density and contribution of DSFGs to the cosmic far-infrared can be a powerful way to validate galaxy formation and evolution models.

(3) The ALMACAL data provide the opportunity to study absorption lines due to galaxies along the line of sight of the calibrators. \cite{klitschALMACALVIMolecular2019} reported several galactic absorption lines using 749 calibrators, but no intervening extragalactic molecular absorber was detected.
They also used the cosmological hydrodynamical simulation IllustrisTNG \citep{naimanFirstResultsIllustrisTNG2018, pillepichFirstResultsIllustrisTNG2018, springelFirstResultsIllustrisTNG2018, marinacciFirstResultsIllustrisTNG2018} to obtain new upper bounds on the molecular gas mass density. Their results are consistent with an increasing depletion of molecular gas in the present Universe compared to redshift $z\sim 2$. In a subsequent study, \cite{klitschALMACALConstraintsMolecular2023} presented the first constraints on the molecular gas coverage fraction in the circumgalactic medium of low-redshift galaxies using estimates of CO column densities along the line of sight of quasars with intervening galaxies.

(4) As ALMACAL observations are repeated over the years, they allow for multi-year follow-up of the AGN variability. 
\cite{bonatoALMACALIVCatalogue2018a} examined 754 calibrator data using Bands 3 and 6 in the time domain space, identifying most of them as blazars based on their flat spectrum and low-frequency spectral index. 
They constructed the light curve of the blazars and found that the median variability index increases steadily with increasing source-to-time lag from 100 to 800 d. 
\cite{husemannJetdrivenGalaxyscaleGas2019} then studied the morphology and kinematics of the gas surrounding the calibrators. They detected a CO($1-0$) emission arc structure around the AGN of the quasar 3C 273. This arc morphology of the molecular gas is completely different from that of the ionised gas. This raises the question of whether the molecular gas is bound to a stellar overdensity formed from a recent galaxy interaction or is currently forming in situ due to a density wave and increased ambient pressure caused by the expanding outflow, as predicted in simulations for luminous AGNs \citep{mukherjeeRelativisticJetFeedback2018}.
\cite{komugiDetectionExtendedMillimeter2022} detected extended mm emission associated with the host galaxy of a prototypical radio-loud quasar to investigate the QSO-host interstellar medium (ISM) interaction.

\begin{figure}
\centering
    \includegraphics[width=1\columnwidth]{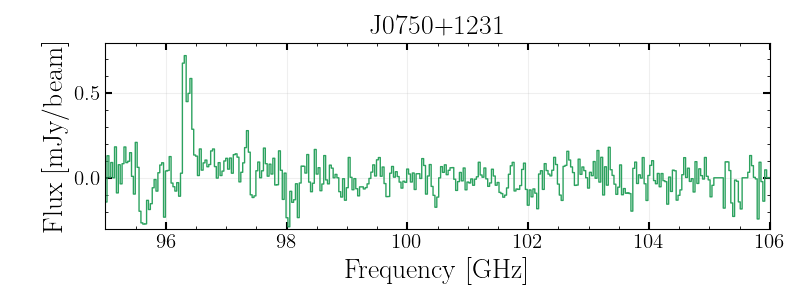}
    \includegraphics[width=0.53\columnwidth]{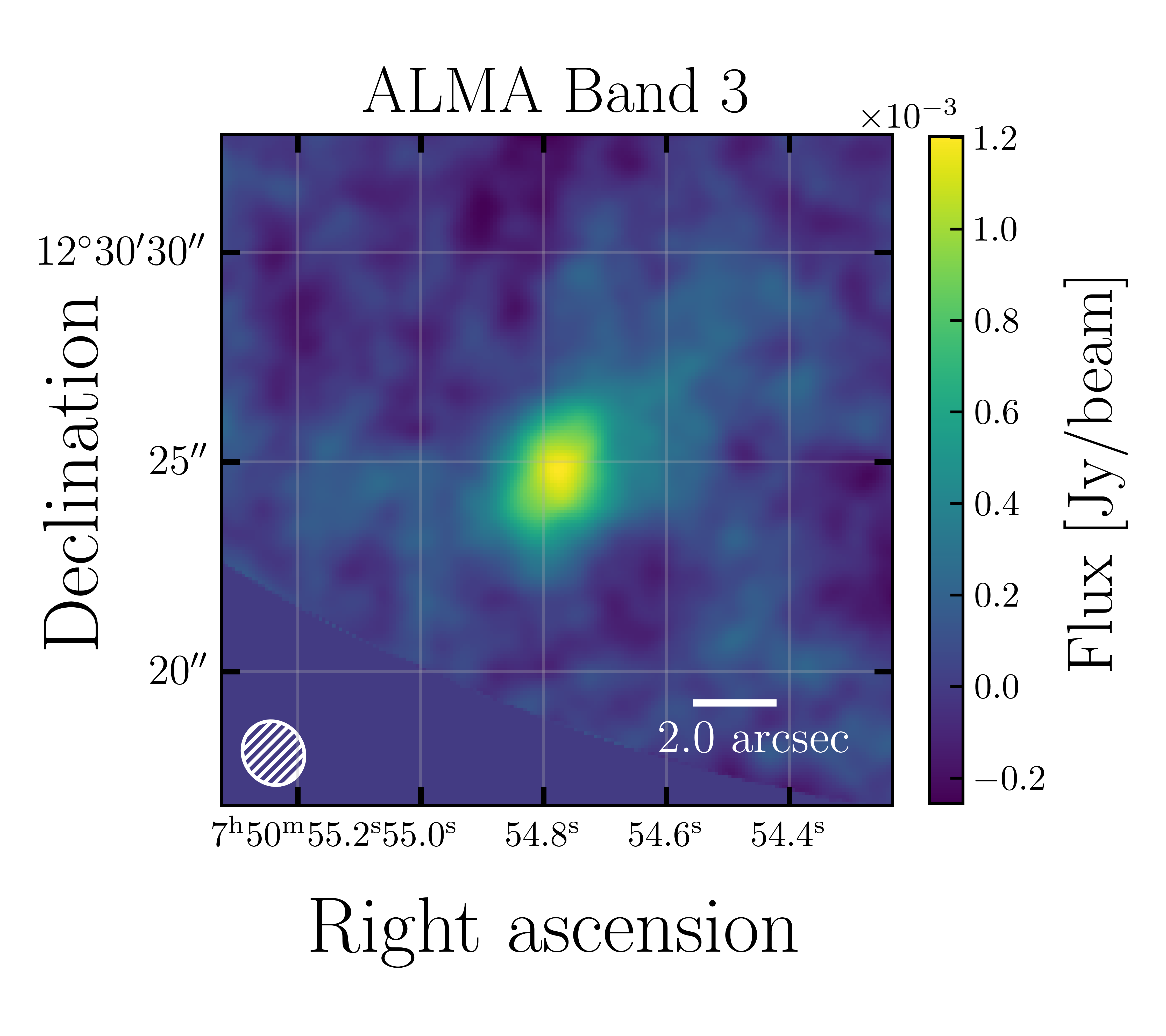} 
    \includegraphics[width=0.46\columnwidth]{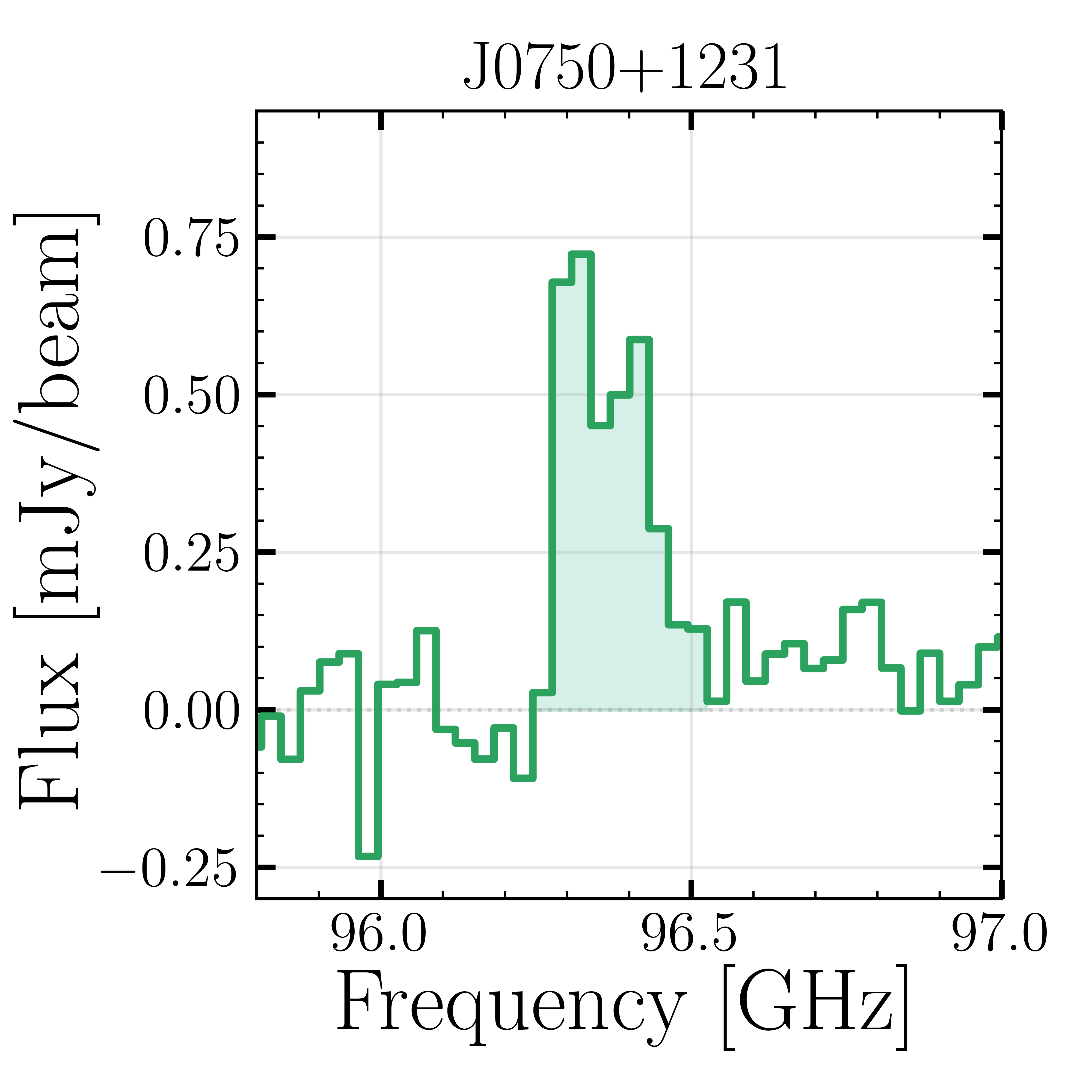}
    \caption{Example of an ALMACAL$-22$ data cube from the updated sample with a prominent emission line found in the calibrator field. The top panel shows the full spectral coverage of the region of the data cube where the emission line was found. The lower left panel displays a region of the continuum map of the calibrator field in Band 3, centred on the position of the emission line. The lower right panel shows the emission line, where the shaded region represents the line width reaching a S/N $\sim9$.}
    \label{fig:science_cases}
\end{figure}

\begin{figure}
\centering
    \includegraphics[width=0.49\columnwidth]{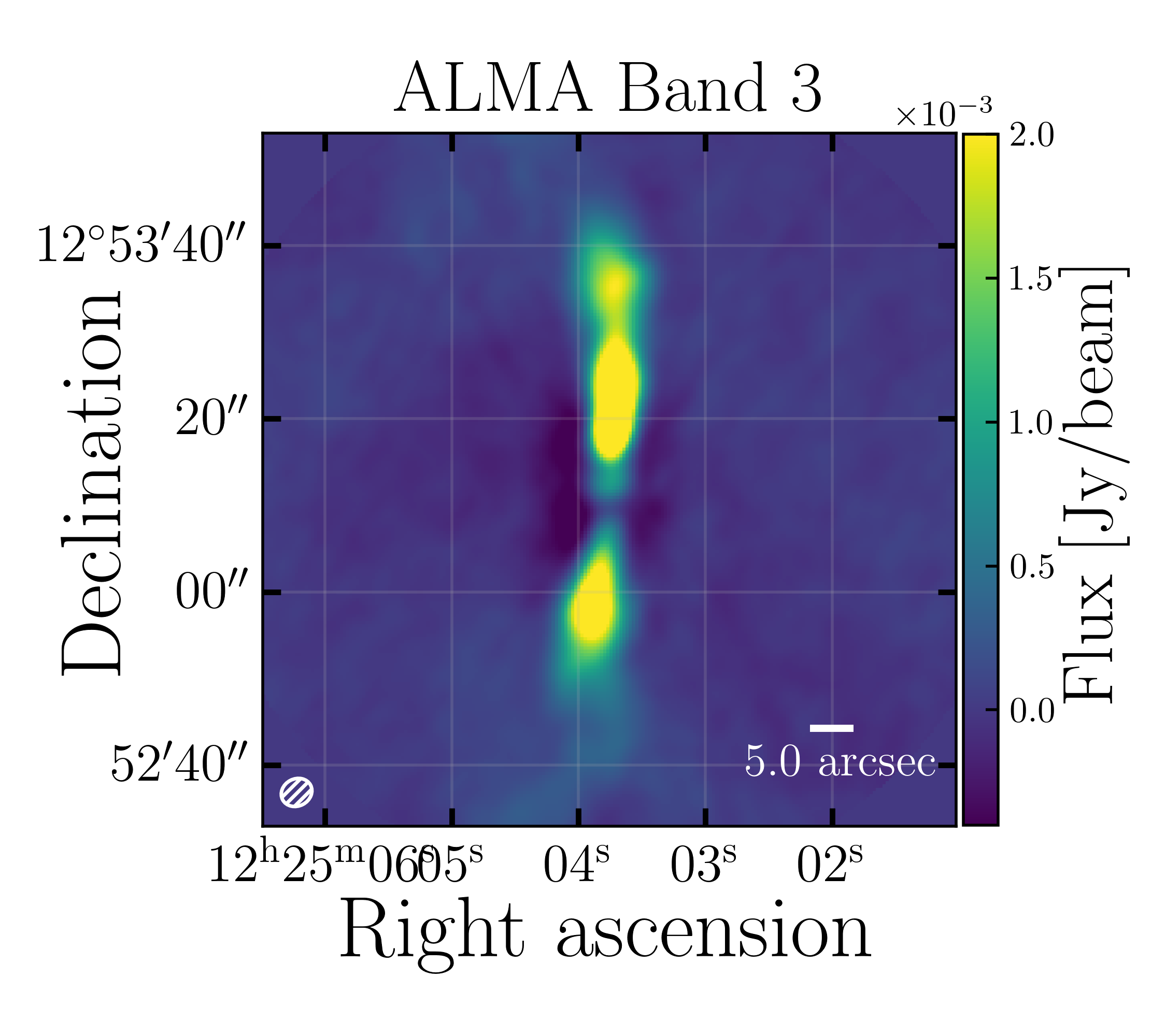}
    \includegraphics[width=0.49\columnwidth]{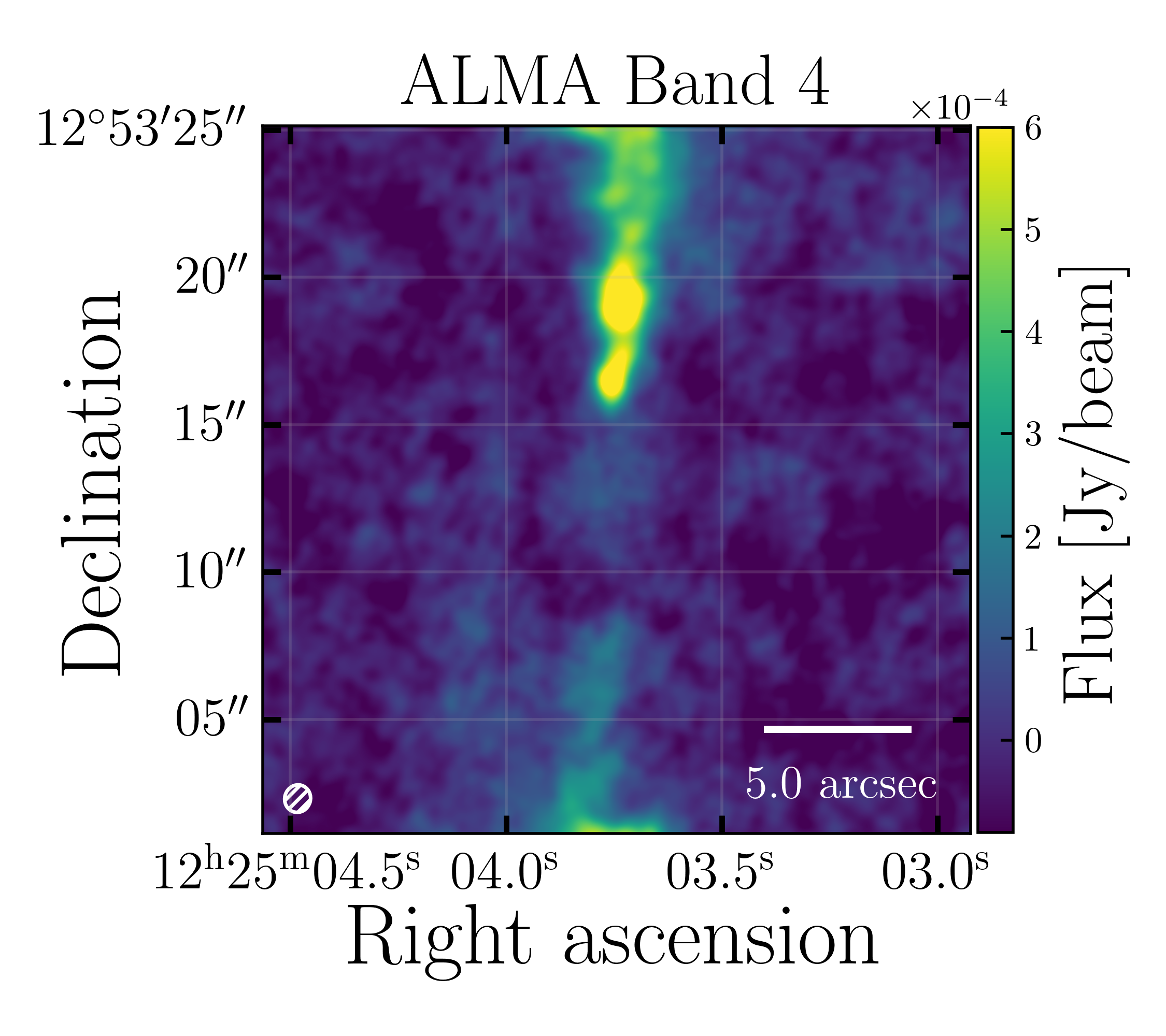} \includegraphics[width=0.49\columnwidth]{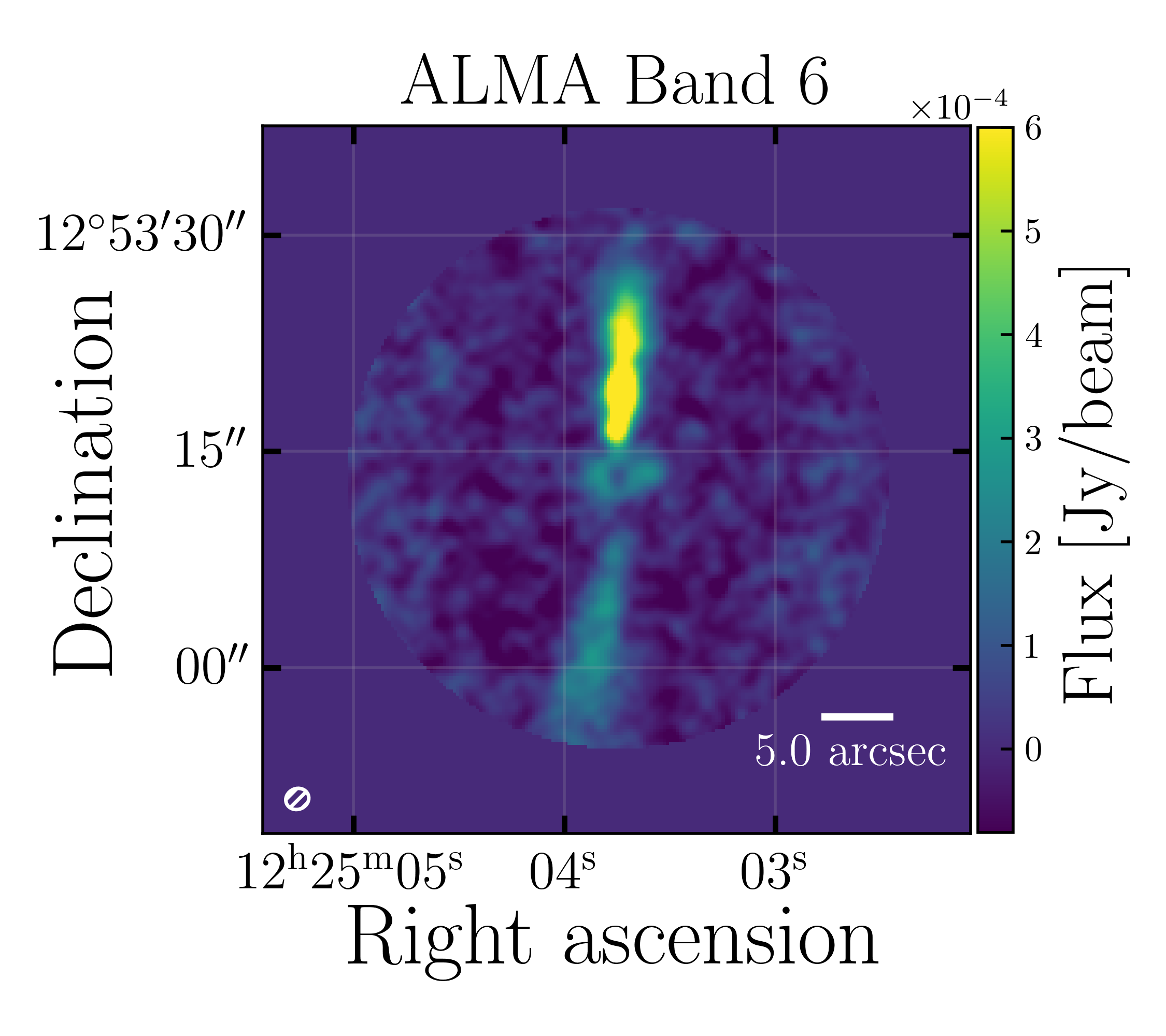}
    \caption{Example of three collapsed data cubes for the calibrator field J1225+1253, where jet emission emanates from the central quasar. ALMA Bands 3, 4, and 6 are shown in the top left, top right, and bottom panels. The central quasar in the centre has previously been subtracted in the {uv} plane.}
    \label{fig:jets}
\end{figure}

The new ALMACAL$-22$ dataset, comprising 1047 calibrator fields, promises significant advancements in our current understanding of these science cases. 
First, the increased calibrator fields will alleviate the field-to-field variance when determining molecular gas evolution. 
The volume probed by each CO transition will increase by a factor of $\sim10$ compared to previous surveys, reducing the uncertainties and expanding at different redshift bins. 
Figure \ref{fig:science_cases} shows an example of a blind search across the calibrator fields, where a prominent emission line was detected.
Second, by adding 46 calibrator fields to the last search for DSFGs \citep{chenALMACALIXMultiband2022} and new observations of the same fields, it will be possible to achieve deeper sensitivity levels, which could improve the estimates of the number counts by detecting the faintest systems in the Universe. 
Thirdly, there are over 300 calibrator fields where the CO absorption lines have yet to be searched for. This updated version of the ALMACAL survey provides a significantly larger dataset than the data analysed in \cite{klitschALMACALAbsorptionselectedGalaxies2019}. 
Finally, ALMACAL$-22$ will allow us to study the variability of specific AGN types such as blazars, including both BL Lacs in the nearby Universe and FSRQs at larger distances.
Calibrators are usually chosen to be point-like, but some have shown extended large-scale jet structures. Figure \ref{fig:jets} illustrates the same calibration field showing extended structures in ALMA Bands 3, 4, and 6. 
Studying the effect of these powerful jets on star formation could help us to understand the high-energy processes better.

\section{Conclusion} \label{sec:conclusions}

ALMACAL$-22$ is a large survey of ALMA calibrator observations, collecting over 30 TB of data and covering 1047 calibrator fields across the southern sky. We have presented all the calibrator data from science projects taken until May 2022, accumulating over 1000 square arcmin and more than 2000 hours of observing time. Here, we provide the characteristics of the survey.

We have presented the selection of a pruned sample, a subset of the highest-quality data.
We outline the data processing details from Band 3 to Band 9 to obtain data cubes as final products.
The pruned sample contains 401 calibrator fields and 1508 data cubes.
We provide an overall review of the main properties of both the full and pruned samples, including the spatial distribution, spatial resolution, integration time, and redshift of the calibrator sources.

In a forthcoming paper, we will revisit the pruned sample to investigate serendipitous detections of the CO emission lines in the calibrator fields. 
In this way, this survey will provide clues to the evolution of molecular gas in the Universe through an untargeted approach.
As ALMACAL$-22$ is one of the largest surveys to date, it will allow us to make statistical estimates that are less sensitive to potential cosmic variance effects.

Overall, ALMACAL is an ever-growing project, as every scientific project requires calibrator data, and the size of the dataset will continue to grow over the years. 
A diverse range of catalogues will be established, including the redshift catalogue, extended jets catalogue, and confirmed molecular line catalogue, showcasing the valuable insights ALMACAL$-22$ will contribute to the scientific community.

\begin{acknowledgements}
      This research was supported by the International Space Science Institute (ISSI) in Bern, through ISSI International Team project \#564 (The Cosmic Baryon Cycle from Space).
\end{acknowledgements}

\section*{Data Availability}
The ALMA Calibrator Source Catalogue is available \href{https://almascience.eso.org/alma-data/calibrator-catalogue}{here}.
ALMACAL raw data and processed data cubes are available upon reasonable request. Please contact \href{almacal@eso.org}{almacal@eso.org}.

  \bibliographystyle{aa} 
  \bibliography{main_corr} 

\end{document}